\documentclass[lettersize,journal]{IEEEtran}
\newtheorem{theorem}{Theorem}

\newtheorem{lemma}{Lemma}
\newtheorem{definition}{Definition}
\newtheorem{remark}{Remark}
\newtheorem{example}{Example}

\usepackage{graphicx}
\usepackage{subfigure}
\usepackage{array}
\usepackage{amsmath}
\usepackage{amssymb}
\usepackage{mathtools}        %数学符号宏包
\usepackage{latexsym}
\usepackage{mathrsfs}
\usepackage{cite}
\usepackage{enumerate} %{enumitem}
\usepackage{float}
\usepackage{amsfonts} % for pdf, bitmapped graphics files
\usepackage{epsfig} % for postscript graphics files
\usepackage{epstopdf}
\usepackage[dvipsnames]{xcolor}
\usepackage{tikz}\newcommand*{\circled}[1]{\lower.7ex\hbox{\tikz\draw (0pt, 0pt)%
		circle (.4em) node {\makebox[1em][c]{\small #1}};}}
\usepackage{color}
\usepackage[scaled]{helvet}
\begin{document}
\title{Prescribed-time Control for Linear Systems in  Canonical Form   Via   Nonlinear Feedback}
	\author{Hefu Ye,   and Yongduan Song$^*$, \IEEEmembership{Fellow, IEEE}
	\thanks{This work was supported by the National Natural Science Foundation
		of China under grant (No.61991400, No.61991403, No.61860206008, and No.61933012). (Corresponding Author: Yongduan Song.)}	
	\thanks{H. F. Ye   is with Chongqing Key Laboratory of Autonomous Systems, Institute of Artificial Intelligence, School of Automation, Chongqing University, Chongqing 400044, China, and also with Star Institute of Intelligent Systems (SIIS), Chongqing 400044, China.   (e-mail:    yehefu@cqu.edu.cn).}
	\thanks{Y. D. Song is with Chongqing Key Laboratory of Autonomous Systems, Institute of Artificial Intelligence, School of Automation, Chongqing University, Chongqing 400044, China. 		(e-mail: ydsong@cqu.edu.cn).}
}

\markboth{IEEE Transactions on Systems, Man, and Cybernetics: Systems, Submitted, Nov. 2021.}%
{Ye} 

\maketitle

	\begin{abstract} 
For systems in canonical form with nonvanishing   uncertainties/disturbances, this work presents an approach to  full state regulation within prescribed time irrespective of initial conditions. By introducing the smooth hyperbolic-tangent-like function, a nonlinear and time-varying state feedback control scheme is constructed, which is further extended to address output feedback based prescribed-time regulation by invoking the prescribed-time observer, all are applicable over the entire operational time zone.	As an alternative to full state regulation within user-assignable time interval,  the proposed method analytically bridges the divide between linear	and nonlinear feedback based prescribed-time control, and is able to achieve asymptotic stability, exponential stability and prescribed-time stability with a unified control structure.
\end{abstract}


\section{Introduction}
\label{sec:introduction}
\IEEEPARstart{F}{inite}-time convergence is highly desirable in many real-world automation applications, where the ultimate control goals are to be realized within finite time rather than infinite time, \textit{e.g.,} auto parts assembling, spacecraft rendezvous and docking \cite{Zhou 2020,Zhao2019-SMC}, and proportional navigation guidance \cite{Zarchan(2012)}, etc. Various  approaches to finite-time convergence have been reported in literature, including: finite-time control, fixed-time control, time-synchronized control, predefined-time control and prescribed-time control.  
The prototype of the finite-time Lyapunov theory originates from $\dot V(x)+ kV^{a}(x)\leq0$ where $V(x)$ is a positive definite function, and $a\in(0,1),~k\in \mathbf{R}^{+}$.  

As an effort to achieve  finite-time stabilization for high-order systems, the homogeneous method, terminal sliding mode method and adding a power integrator method are successively proposed (see, for instance, \cite{Bhat 2005,Levant2003,Levant2005,Sun 2017,Chen 2013,Amato2001,Bhat 1998,Zhao2018}), which  greatly promote the development of finite-time control theory.  Since  the convergence (settling)  time therein depends on initial conditions and design parameters,  the notion of  fixed-time control  is then introduced in \cite{Poly 15}  and  \cite{Poly 12}, where the   fractional-order plus odd-order feedback is used, leading to  different closed-loop system dynamics, so that the upper boundary of the convergence time can be estimated  without using  initial conditions.
However, neither finite-time control nor fixed-time control can actually achieve state regulation within one unified time.    The time-synchronized control scheme proposed in  \cite{Ge2021} and \cite{Li2021},  based on the norm-normalized sign function,  is shown to be able to achieve output regulation simultaneously for different initial conditions with a unified control law.

To further alleviate the dependence of the settling time on design parameters, a predefined-time  approach is exploited to estimate the upper bound of the convergence time in \cite{Sanchez-Torres 2015} and \cite{Sanchez-Torres 2018} by multiplying exponential  signals on the basis of fractional power feedback signals. 
Recently, the notion of prescribed-time control is proposed in \cite{Song 17}, which allows the user to assign the convergence time at will and irrespective of initial conditions or any other design parameter, thus offers a clear advantage over those that do not.  {With this concept},   three different approaches have been developed, namely, states transformation approach, temporal scale transformation  approach, and parametric Lyapunov equation based approach (\textit{e.g.,} \cite{Song 17,Krishnamurthy 20a,Krishnamurthy2019,Zhou 20,Song 19,Kanzhen1,Ye2021}). Based on states transformation approach, the distributed consensus control algorithms are studied for multi-agent systems in  \cite{Song 2018,wang-prescribed-time,wang-zeroerror}, and a prescribed-time observer   based output feedback algorithm is elegantly established for linear systems in  \cite{Holloway 19b}.  Subsequent works   further consider more complex systems, such as stochastic nonlinear systems    \cite{Liwuquan2021} and LTI systems with input delay \cite{France2021}.  In addition, by using temporal scale transformation, a triangularly stable  controller is proposed for  the  perturbed system  in \cite{Shakouri21},   a  dynamic high-gain   feedback   algorithm is established for strict-feedback-like systems in \cite{Krishnamurthy2019}, and some distributed algorithms are developed for multi-agent systems in \cite{Kanzhen2,Kanzhen3,Yucelen3}.  Based upon parametric Lyapunov equation, a finite-time controller and a prescribed-time controller  are studied for linear systems in \cite{Zhou 2020} and \cite{Zhou 20},  and then generalized to  nonlinear systems in \cite{zhoubin}.

Theoretically inclined,   prescribed-time control systems, under some generic design
conditions, are capable of tolerating large parametric, structural
and parameterizable disturbance uncertainties on the finite time interval, to ensure desired control performance, in addition to system stability. This property comes from a time-varying function which goes to $\infty$ as $t$ tends to the prescribed time.   
Different from  \cite{Song 17,Song 2018,Song 19}, where the time-varying function is used to scale the coordinate transformations, this paper only introduces the time-varying function into the virtual/actual controller. The advantage of this approach is that a simpler controller results and the control effort is reduced.
In addition, to obtain far superior transient performance, we choose a new feedback scheme that the regular feedback signal can be  reconstructed into some suitable forms by a nonlinear mechanism  with high design degrees of freedom.\footnote{Some early literature exploit similar ideas to improved control performance. For example, in classical proportional-integral-differential (PID) control, the feedback signal $x$ is processed by proportion, integral or differential to construct the classical PI, PD or PID control.
}

Motivated by the above discussions,  this paper  revisits the prescribed-time control of high-order systems  via a novel nonlinear feedback approach.  In Section \ref{section2},   a  useful hyperbolic-tangent-like function and a novel lemma are presented. In Section \ref{section3}, we study the prescribed-time control for certain LTI systems by using a nonlinear and time-varying feedback, both full state feedback and partial state feedback are considered. Section \ref{section4} gives a extend prescribed-time control algorithm for uncertain LTI systems. Section \ref{section5} concludes this paper. The main contributions of this paper are as follows:
\begin{itemize}
\item 	Both full state and partial state feedback controller are designed to achieve state regulation within prescribed time irrespective of initial conditions and any other design parameter. A non-stop running implementation method with ISS property is proposed for the first time. 
	\item   %far superior transient performance without an increase in control effort.  
	  Unlike most existing solutions that usually use the regular (direct) state feedback, this paper proposes to use the ``reshaped” feedback states through the hyperbolic-tangent-like function, so as to establish a nonlinear and time-varying feedback control strategy capable of addressing asymptotic, exponential and prescribed-time control uniformly under certain conditions.  
	\item For high-order systems with   non-parametric uncertainties/disturbances, we propose a prescribed-time sliding mode control scheme, melting attractive stability and robustness features at the transition and steady-state stages.
\end{itemize}

% By constructing a hyperbolic-tangent-like function to reshape the feedback signals,   we build a unified control scheme capable of achieving asymptotic, exponential and prescribed-time regulation (with the related design parameters chosen appropriately). Due to the nature of the nonlinear and time-varying feedback, excessively large  initial control effort is avoided even for  large initial conditions (Sections 2-3). The control is designed collectively with one step and one Lyapunov function. By combining the prescribed-time observer ({\color{RoyalBlue}\citealp{Holloway 19a}}), the proposed control is also able to achieve output feedback based state regulation within prescribed time irrespective of initial conditions and any other design parameter  (Section 4). For  high-order systems with matched non-parametric uncertainties/disturbances, we propose a prescribed-time sliding mode control scheme,    melting attractive stability and robustness features at the transition and steady-state stages (Section 5).  All established results are valid over infinite time interval. 
\textit{Notations:} 
$\mathbf{R}$ is the set of nonnegative real numbers, $\mathbf{R}^+=\{x\in \mathbf{R}:x>0\}$.
For non zero integers $m$ and $n$, let $0_{m\times n}$ be the
$(m, n)$-matrix with zero entries, and
$J_n=((0_{(n-1)\times1},I_{n-1})^{\top},0_{n\times1})^{\top}$ 	 and $\mathcal{L}_n(\mathbf{a})=\left((0_{n\times(n-1)}),\mathbf{a}\right)^{\top}$ where $\mathbf{a}=(a_1,\cdots,a_{n})^{\top}$.
%$D^*f(t)$ denotes Dini derivative of $f(t)$.
$\ell_{\infty}[0,t_p)$ denotes $\ell_{\infty}$ on $[0,t_p)$.   Denote by $\mathcal{K}$ the set of class $\mathcal{K}$-functions  and denote by $\mathcal{KL}$ the set of class $\mathcal{KL}$-functions (see Section 4.4 in  \cite{Khalil(2002)}). For any vector $\mathbf{z}$, we use $\mathbf{z}^{\top}$ and $\|\mathbf{z}\|$ to denote its transpose and Euclidean norm respectively. $\lim_{t\rightarrow T}f(\cdot)$ denotes the limit of $f(\cdot)$ as $t\rightarrow T$. We denote by $\bullet^{(q)}(q=0,\cdots,n)$ the $q$-th derivative of $\bullet$, and denote by $\bullet^{q}$ the $q$-th power of $\bullet$.

\section{Preliminaries}\label{section2}
\subsection{Problem Statement}
{We  restrict our analysis to the following system in canonical form with uncertainties/disturbances }
\begin{equation}\label{SISOsystem}
\left\{\begin{array}{ll}
\begin{aligned}
\dot{\mathbf{x}}(t)&=A{\mathbf{x}}(t)+Bu(t)+D(\mathbf{x},t)\\
y(t)&=C{\mathbf{x}}(t)
\end{aligned}
\end{array}\right.
\end{equation}
where $A=J_n+\mathcal{L}_n(\mathbf{a})$ is the system matrix,  $B=[0,\cdots,0, 1]^{\top} $, $C=[b_0,b_1,\cdots,b_{n-1}]$ are coefficient vectors, $D=[0,\cdots,0,d(\mathbf{x},t)]^{\top}$ with $d(\mathbf{x},t):\mathbf{R}^n\times [0,\infty)\rightarrow \mathbf{R}$  modeling the unknown nonvanishing uncertainties/disturbances of the system, ${\mathbf{x}}(t)=[x_1,\cdots, x_n]^{\top}\in \mathbf{R}^n $ is the vector of system states, and $u(t)\in\mathbf{R}$ is the control input. $(A,B)$ is controllable and $(A,C)$ is observable. The control objective is to design a feedback  control $u(t)$ to  stabilize (\ref{SISOsystem}) within prescribed-time $t_p$, \textit{i.e.,}  $\mathbf{x}(t) \in\ell_{\infty}[0,t_p)$ and $\lim_{t\rightarrow t_p} \{x_i(t)\}_{i=1}^n=0.$  We are particularly interested in making use of  the feedback information  $x$ through a nonlinear way to construct the control scheme. 

\begin{definition}{\cite{France2021}}
	The origin of the system $\dot{\mathbf{x}}=f(\mathbf{x},t)$ is said to be prescribed-time globally asymptotically stable  (PT-GUAS) if there exist a class $\mathcal{K} \mathcal{L}$ function $\beta$ and a function $\mu:\left[0,  t_p\right) \rightarrow \mathbf{R}^{+}$such that $\mu$ tends to infinity as $t$ goes to $t_p$ and, $\forall t \in\left[0, t_p\right)$
	$$
	\|\mathbf{x}(t)\| \leq \beta\left(\left\|x\left(0\right)\right\|, \mu\left(t\right)\right),
	$$
	where $t_p$ is a time that can be prescribed in the design. 
\end{definition}

\subsection{Hyperbolic-tangent-like Function}
Instead of using $x$ directly, we process the feedback information $x$ through the following hyperbolic-tangent-like function  $h(x):(-\infty,+\infty)$ $\rightarrow $ $\left(-\frac{1}{b},\frac{1}{a}\right)$ as:
\begin{equation}\label{hx}
h(x)\coloneqq\frac{e^{a x}-e^{-bx}}{ae^{ax}+b e^{-b x} },~0<                             b\leq a
\end{equation}
where $a$ and $b$ are design parameters, which becomes  the standard hyperbolic tangent function for $a=b=1$. Such nonlinear feedback exhibits two salient properties.

 \textit{Property 1:} \textit{The function  $h(x)$ is $\mathcal{C^{\infty}}$ on $\mathbf{R}$  and $ h(x)=0$ if and only if $x=0$. Under $0<b\leq a$, the inequality $0\leq|x|h(|x|)\leq xh(x)$ holds.}
 	
	\textit{Proof:} Define a continuous function $F(x)=xh(x)-|x|h(|x|)$. For $\forall x\geq 0$, we have    
	\begin{equation*}
	F(x)= 
	\frac{x(e^{a x}-e^{-b x})}{a e^{a x}+b e^{-b x}}-\frac{|x|(e^{a |x|}-e^{-b |x|})}{a e^{a |x|}+b e^{-b |x|}}\equiv 0.
	\end{equation*}    	
	For $ \forall x<0$, it follows that   
	\begin{equation*}
	\begin{aligned}
	F(x)=&
	\frac{x(e^{a x}-e^{-b x})}{ae^{a x}+b e^{-b x}}-\frac{x(e^{b x}-e^{-a x})}{b e^{b x}+a e^{-a x}}\\
	=&\frac{x(b-a)\left(e^{(a+b)x}+e^{-(a+b)x}-2\right)}{\left(a e^{a x}+b e^{-b x}\right)\left(b e^{b x}+a e^{-a x}\right)}\geq 0.
	\end{aligned}
	\end{equation*}  
	Thus, under $0<b\leq a$, the inequality $F(x)\geq0$ holds for $\forall x\in(-\infty,+\infty)$, implying that $0\leq|x|h(|x|)\leq xh(x)$  for $\forall x\in(-\infty,+\infty)$.   $\hfill\blacksquare$
 
\textit{Property 2:} \textit{Function $h(x)$ is strictly monotonically increasing with respect to (\textit{w.r.t.}) $x$, its upper and lower bounds are $1/a$ and $-1/b$, respectively. By selecting different design parameters $a$ and $b$,  various functions can be obtained from $h(x)$,  In particular,  if choosing $a$ and $b$ small enough, then $h(x) = x$.} 
 
	\textit{Proof:} The upper and lower bounds of $h(x)$ are
	\begin{equation*} 
	\begin{aligned}
	\lim_{x\rightarrow +\infty}h(x)&=\lim_{x\rightarrow +\infty}\frac{e^{a x}-e^{-b x}}{a e^{a x}+be^{-bx} }=\frac{1}{a},~~~\alpha>0\\
	\lim_{x\rightarrow -\infty}h(x)&=\lim_{x\rightarrow -\infty}\frac{e^{a x}-e^{-bx}}{a e^{a x}+b e^{-b x} }=-\frac{1}{b},~b>0.~~~~~~~~~
	\end{aligned}
	\end{equation*}
	When $a$ and $b$ are sufficiently small,  for $~x\in(-\infty,+\infty)$, by using \textit{L'H$\hat{\text{o}}$pital's Rule}, we have
	\begin{equation*} \label{5}
	\begin{aligned}
	\mathop {\lim }_{a\rightarrow 0\atop b\rightarrow 0 }\frac{e^{a x}-e^{-b x}}{a e^{a x}+b e^{-b x} } &=\mathop {\lim }_{a\rightarrow 0\atop b\rightarrow 0 }\frac{xe^{a x}+xe^{-b x}}{e^{a x}+a xe^{a x}+e^{-b x}-b xe^{-b x} }\\&=x 
	\end{aligned}
	\end{equation*}
	implying that the expanded/compressed signal $h(x)$ reverses back to the regular signal $x$.   $\hfill\blacksquare$

%The function $\mu(t,t_p)$ is $\mathcal{C^{\infty}}$ on $[0,t_p)$, and $h(x)$ is $\mathcal{C^{\infty}}$ on $\mathbf{R}$, where these $\mathcal{C^{\infty}}$ smoothness feature will be used for later control design. In addition, Property 2 indicates that  the design parameters $a$ and $b$  influence the characteristics of feedback signals.

\begin{remark} The classical finite-time control adopts fractional  power of $x$ (\textit{i.e.,} $x^{1/3}$) to expand  the feedback signal $x$ on $x \in[-1, 1]$,	and compress  $x$ on $x\in(-\infty,-1)\cup(-1,+\infty)$; the fixed-time control uses an additional nonlinear damping term $(e.g., x^3)$ on the basis of the original feedback signal to expand  the feedback signal on $(-\infty,+\infty)$.  Consequentially, different feedback signals cause different convergence properties,  which is mainly reflected in the relationship between settling time and initial conditions. 	
	Since the hyperbolic-tangent-like function $h(x)$ can expand  or compress the feedback signal with different $a$ and $b$, it provides control design extra flexibility and degree of freedom. 
	In addition,  the right-hand side of (\ref{hx}) remains bounded within $(-1/b,1/a)$ even if $|x|$ grows large, this special property  makes $h(x)$ perfectly suitable to function as the core part of the controller, and our motivation for this work partly stems from such appealing features of $h(x)$. Fig. 1 illustrates  $h(x)$ with different $a$ and $b$, and the  fractional power feedback signals in terms of $x$. 
\end{remark}
\begin{figure}[htbp]\label{fig1}
	\vspace{-0.1cm}
	\begin{center}
		\includegraphics[height=4.6cm]{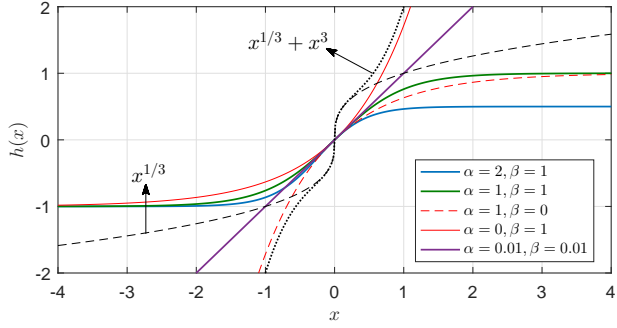}
		\caption{ {Schematic figure of $h(x)=\frac{e^{a x}-e^{-b x}}{a e^{a x}+b e^{-b x}}$ and the trajectories of  $x^{1/3}+x^3$ and $x^{1/3}$.} }
	\end{center}
\end{figure}

 The interesting feature behind this nonlinear feedback is that it includes regular (direct) feedback of $x$ as a special case,  and it allows the linear regular feedback control and nonlinear feedback control to be unified through such function,    providing a variety of  ways to making use of $x$ for control development.
By using properties of  the hyperbolic-tangent-like function, we establish the following lemma, which is crucial to our later technical development. 
\begin{lemma}
	Consider the functions $\mu(t)=1/(t_p-t)$ and $h(x)$ be given
	as in   (\ref{hx}). For $t\in[0,t_p)$, if a positive continuously differentiable	function  satisfies  
	\begin{equation}\label{lemma11}
	\dot{V}(t)\leq -k\mu(t) h(V),~k>1
	\end{equation} 
	where $\dot{V}=-k\mu(t)h(V)$ if and only if $V=0$,  then we have $V(t)\leq \beta(V_0,\mu(t))$ and $\beta$ being of class-$\mathcal{KL}$. In particular, it holds
	\begin{equation*}
	\lim_{t\rightarrow t_p}V(t)=0,~\lim_{t\rightarrow t_p}\dot{V}(t)=0.
	\end{equation*} 
\end{lemma}

\textit{Proof:} Consider the  analytical expression of (\ref{lemma11}):
\begin{equation}\label{Lemma11}
\dot{V} \leq \frac{-k}{t_p-t}\frac{e^{a V}-e^{-b V}}{a e^{a V}+b e^{-b V}},
\end{equation}
Let $V_*= e^{aV}-e^{-bV} $, where ${V_*}_0=e^{aV(0)}-e^{-bV(0)}$. Then we have $V_*\geq0$, and
\begin{equation}
\begin{aligned}
\dot V_*\leq& \left(ae^{aV}+be^{-bV}\right)\frac{-k}{t_p-t}\frac{e^{a V}-e^{-b V}}{a e^{a V}+b e^{-b V}}\\
\leq&-k\mu V_*
\end{aligned}
\end{equation}
holds for $t\in[0,t_p)$. Hence we derive $V_*\leq \beta({V_*}_0,\mu(t))$ according to Lemma 1 in \cite{Song 17}, namely $V_*\in\ell_{\infty}[0,t_p)$ and $\lim_{t\rightarrow t_p}V_*=0$. The same result can be established for  $(e^{aV}-e^{-bV})$. 
It follows from the fact $V$ tends to $+\infty$ ``slower" than $(e^{aV}-e^{-bV})$ as $V\rightarrow +\infty$  that $V(t)\in\ell_{\infty}[0,t_p)$ and $\lim_{t\rightarrow t_p}V(t)=0$.
In addition, the inequality (\ref{Lemma11}) can be transformed into the following form:
\begin{equation} \label{x}
V\leq\frac{1}{a}\ln\left(C_1(t_p-t)^k+e^{-b V}\right),~ t\in[0,t_p) 
\end{equation}
where $C_1=\left(e^{a V_0}-e^{-b V_0}\right)/{t_p^k}$ is the integral constant. 
In fact, one can easily verify the following calculations
	\begin{equation*}
	\begin{array}{|l|c|}
	\hline \small{\text{(6)}}\Rightarrow ~e^{a V}-e^{-b V}\leq C_1(t_p-t)^k; & ~\small{\text{(6.1)}}\\
	~~~ \Rightarrow \frac{e^{a V}-e^{-b V}}{t_p-t}\leq C_1(t_p-t)^{k-1}; & ~\small{\text{(6.2)}}\\
	\small{\text{(6.1)}}\Rightarrow ~\left(a e^{a V}+b e^{-b V}\right)\dot V\leq-kC_1(t_p-t)^{k-1};~& ~\small{\text{(6.3)}}\\
	\small{\text{(6.3)}}\Rightarrow ~\dot V\leq\frac{-kC_1(t_p-t)^{k-1}}{a e^{a V}+b e^{-b V}}\leq \frac{-k}{t_p-t}\frac{e^{a V}-e^{-b V}}{a e^{a V}+b e^{-b V}}. ~& ~\small{\text{(6.4)}}\\
	\hline
	\end{array}
	\end{equation*} 
	Furthermore,  from (6.3) we have $(a+b)\dot V\leq 0$ and 
	\begin{equation}\label{DV}
	\dot V\leq \frac{-kC_1(t_p-t)^{k-1}}{a e^{a V}+b e^{-b V}},~\dot V(0)=\frac{-kC_1 t_p^{k-1}}{a e^{a V_0}+b e^{-b V_0}}.
	\end{equation}
	Indeed, notice that  $\dot{V}$ is a continuous function and $\dot V=-k\mu h(V)$ for $V=0$, implying that $\dot V\in\ell_{\infty}[0,t_p) $ and $\dot V(t)\rightarrow 0 $ as $t\rightarrow t_p$.  This completes the proof. $\hfill\blacksquare$

As discussed earlier, when $a$ and $b$ are sufficiently small, we have $\lim_{a,b\rightarrow 0}h(V)=V$. Lemma 1 is therefore equivalent to the Corollary 1 in \cite{Song 17}. 

\section{Prescribed-time control for linear systems in canonical form  without uncertainties}\label{section3}
Motivated by the appealing features of the nonlinear scaling function $h(x)$, we now discuss how to introduce it into the prescribed-time control design of $n$-th order systems (\ref{SISOsystem}).  We first design prescribed-time control schemes  using nonlinear and time-varying  full state feedback  and partial state feedback to achieve full state regulation for system (\ref{SISOsystem}) without uncertainties/disturbances (\textit{i.e.,} $d(\mathbf{x},t)\equiv {0}$), then we extend the control scheme to cope with nonvanishing uncertainties/disturbances in the system.  
\subsection{Prescribed-time State Feedback  Controller}
By using the time-varying scaling function and the hyperbolic-tangent-like function as introduced in Section 2, we construct the vectors as
\begin{equation}
\begin{aligned}
H(\mathbf{z})&= [h(z_1),  \cdots,h(z_n)]^{\top} \\
\Gamma(\mathbf{z})&=k\mu^rH(\mathbf{z}) 
\end{aligned}
\end{equation}  
where  $\mu=1/(t_p-t)$ and $h(\cdot)$ is defined   in   (\ref{hx}).  In addition, we introduce the following two auxiliary vectors 
\begin{equation}\label{auxiliaryvector}
\begin{aligned}
\mathbf{z}&={\mathbf{x}}+J_n^{\top}\Phi\\
\Phi&=J_n^{\top}(\dot{\Phi}+{\mathbf{z}})+\Gamma(\mathbf{z})
\end{aligned}
\end{equation}
where $\mathbf{z}=[z_1,\cdots,z_n]^{\top}\in \mathbf{R}^n$, $\Phi=[\phi_1,\cdots,\phi_n]^{\top}\in \mathbf{R}^n$. 
Note that $J_n^{\top}$ is lower triangular, thus both   $\mathbf{z}$ and $\Phi$   can be easily calculated recursively (see (\ref{228}) for specific example of computing $z_i$ and $\phi_i$). 
It is interesting to see   that  the convergence property of the closed-loop system only depends on the parameters $a$, $b$,   $k$ and $r$ in vector  $\Gamma(\mathbf{z})$. 
\begin{theorem}
	Consider system (\ref{SISOsystem})  with $d(\mathbf{x},t)\equiv {0}$ and the state feedback control law, 
	\begin{equation}\label{u}
	\begin{aligned}
	u=- {B^{\top}} (\mathcal{L}_n(\mathbf{a})\mathbf{x}+{\Phi}),
	\end{aligned}
	\end{equation}
	then all closed-loop signals are bounded  and the origin of the closed-loop  system (\ref{SISOsystem})  is:   
	\begin{itemize}
		\item[$1)$] \textbf{Globally uniformly asymptotically  stable (GUAS)},  that is, $\|x(t)\|\leq \beta(\|\mathbf{x}(0)\|,t)$, if the controller parameters are selected as  $k>0$ and $r=0$. In addition,    exponential output regulation is achieved if $a$ and $b$ are chosen sufficiently small. 
		\item[$2)$] {\textbf{Prescribed-time globally uniformly asymptotically stable (PT-GUAS)}, that is, $\|x(t)\|\leq \beta(\|\mathbf{x}(0)\|,\mu(t))$, if the controller parameters are selected as $k>n$ and $r=1$.} %Furthermore, if  $u(t)$ is set as $u(t)\equiv 0,~\forall t\in [t_p,+\infty)$, then $\{x_i(t)\}_{i=1}^n$ remain zero for all $t\geq t_p$.
	\end{itemize}

\end{theorem}

\textit{Proof:} 
{By the definition of $J_n$, $B$ and $\mathcal{L}_n(\mathbf{a})$, it is readily verified that $J_nJ_n^{\top}+BB^{\top}=I_n$ and {$BB^{\top}\mathcal{L}_n(\mathbf{a})=\mathcal{L}_n(\mathbf{a})$}, therefore, it holds that 
	\begin{equation}\label{dz}
	\begin{aligned}
	\dot {\mathbf{z}}&=A{\mathbf{x}}+B u+J_n^{\top}\dot{\Phi}\\
	&=\left(J_n+\mathcal{L}_n(\mathbf{a})\right){\mathbf{z}}-J_nJ_n^{\top}{\Phi}-\mathcal{L}_n(\mathbf{a})J_n^{\top}{\Phi}\\
	&~~~-BB^{\top}\mathcal{L}_n(\mathbf{a}){\mathbf{x}}-BB^{\top}{\Phi}+{\Phi}-\Gamma(\mathbf{z})-J_n^{\top}{\mathbf{z}}\\
	&=\left(J_n+\mathcal{L}_n(\mathbf{a})\right){\mathbf{z}}-{\Phi}-\mathcal{L}_n(\mathbf{a})\left(J_n^{\top}{\Phi}+{\mathbf{x}}\right)\\
	&~~~+{\Phi}-\Gamma(\mathbf{z})-J_n^{\top}{\mathbf{z}}\\
	&=\left(J_n-J_n^{\top}\right){\mathbf{z}}-\Gamma(\mathbf{z}).
	\end{aligned}
	\end{equation}
	\textit{ 1) proof of  GUAS result:}  we prove that the closed-loop  system under the control law (\ref{u}) with $r=0$ (in this case, $\Gamma(\mathbf{z})=kH(\mathbf{z}) $) is GUAS. 
	To this end, we define the error vector between $H(\mathbf{z})$ and $\mathbf{z}(t)$ as 
	\begin{equation}\label{12}
	\mathbf{E}\coloneqq H(\mathbf{z})-\mathbf{z}(t)
	\end{equation} 
	where $\mathbf{E}=[e_1,\cdots,e_n]^{\top}\in \mathbf{R}^n$ is a smooth function  satisfying $\lim_{\alpha,\beta\rightarrow 0}\mathbf{E}=\mathbf{0}$
	and $\mathbf{E}$ is bounded as long as $\mathbf{z}(t)$ is bounded. Consider a positive definite function $V_1=\mathbf{z}^{\top}\mathbf{z}/2$, the time derivative of $V_1$ along (\ref{dz}) is
	\begin{equation}\label{ddV}
	\dot V_1=\mathbf{z}^{\top}\left(J_n-J_n^{\top}\right){\mathbf{z}}-\mathbf{z}^{\top}\Gamma(\mathbf{z})=-k\mathbf{z}^{\top}H(\mathbf{z}) \leq 0
	\end{equation}
	where the fact that  ${\mathbf{z}}^{\top}\left(J_n-J_n^{\top}\right){\mathbf{z}}=0,~\forall {\mathbf{z}}\in\mathbf{R}^n$ is used since $J_n-J_n^{\top}$ is a skew symmetric matrix.  	It follows from (\ref{ddV})  that $\dot V_1=0$ if and only if $\mathbf{z}=\mathbf{0}$, thus the transformed system (\ref{dz}) is asymptotically stable on $[0,+\infty)$, establishing the same to system (\ref{SISOsystem}) according to the converging-input converging-output property of the corresponding auxiliary vectors.

	From (\ref{ddV}), it can be further shown  that
	\begin{equation}\label{dV}
	\begin{aligned}
	\dot V_1&=-k\mathbf{z}^{\top}(\mathbf{E}+\mathbf{z})\leq-k\|\mathbf{z}\|^2+k\|\mathbf{z}\|\|\mathbf{E}\|\\
	&\leq-\frac{k}{2}\|\mathbf{z}\|^2+\frac{k}{2}\Delta^2
	\end{aligned}
	\end{equation} 
	where $\Delta \coloneqq\sup\{\mathbf{\|E\|}\}$. 	By integrating both sides of the inequality (\ref{dV}), we obtain $V_1(t)\leq V_1(0)e^{-kt/2}+\Delta^2(1-e^{-kt/2})$. 
	If we further choose the design parameters $a$ and $b$ in $H(\mathbf{z})$ small enough, one can obtain  $\lim_{a,b\rightarrow 0}\Delta=0$, yielding $V_1(t)\leq V_1(0)e^{-kt/2}$, thus we have $\|\mathbf{z}(t)\|\leq \|\mathbf{z}(0)\| {e^{-kt/2}}$ and the transformed system (\ref{dz}) is  exponentially stable. In addition, it follows from (\ref{auxiliaryvector}) that $x_1(t)=z_1(t)$, therefore exponential output regulation to zero of (\ref{SISOsystem}) can be achieved.  The word ``exponential" actually means that ``near exponential", because the parameters $a$ and $b$ can only be selected as sufficiently small, not zero.

	\textit{2) Proof of PT-GUAS result:} we now show that the closed-loop system under the control law (\ref{u})  with $r=1$ (in this case, $\Gamma(\mathbf{z})=k\mu H(\mathbf{z}) $)  is PT-GUAS. For $t\in[0,t_p),$ there exists a continuous positive function $W(\mathbf{z},t)=\sqrt{\mathbf{z}^{\top}{\mathbf{z}}/n} $ such that $W(\mathbf{z},t)=0$ as $\mathbf{z}=\mathbf{0}$ and 
		\begin{equation} \label{z*}
		W^2=\mathbf{z}^{\top}{\mathbf{z}}/n \Rightarrow W\leq   \max\{|z_1|,\cdots,|z_n|\}\triangleq  {z}^*
		\end{equation}	
		In addition,  its time derivative can be shown as
		\begin{equation}\label{V}
		\begin{aligned}
		\dot W=& \frac{\mathbf{z}^{\top}\dot{\mathbf{z}}}{nW}=\frac{{\mathbf{z}}^{\top}\left(J_n-J_n^{\top}\right){\mathbf{z}}-\mathbf{z}^{\top}\Gamma(\mathbf{z})}{nW}\\
		=&-\frac{ \mathbf{z}^{\top}\Gamma(\mathbf{z})}{nW}=-\frac{k\mu}{nW}\sum_{i=1}^{n}z_ih(z_i) .
		\end{aligned}
		\end{equation}
		Using Property 1 and (\ref{z*}), we have	
		\begin{equation} \label{W}
		\begin{aligned}
		\dot W&\leq-\frac{k\mu}{nW}\sum_{i=1}^{n}|z_i|h(|z_i|) \leq -\frac{k\mu}{nW} {z}^*h( {z}^*)\\
		&\leq -\frac{k\mu}{nW}  {W}  h(W)=- \frac{k\mu}{n} h(W)\leq 0.
		\end{aligned}
		\end{equation}		
		Since we select $k>n$, then $k/n>1$. By virtue of Lemma 1, one can prove that  $W(t)\in\ell_{\infty}[0,t_p)$, $\dot{W}(t)\in\ell_{\infty}[0,t_p)$ and 
		\begin{equation}
		\lim_{t\rightarrow t_p}W(t)=0,~\lim_{t\rightarrow t_p}\dot W(t)=0.
		\end{equation}  
		Hence the closed-loop signals $\{z_i\}_{i=1}^n\in\ell_{\infty}[0,t_p)$ and $\{\dot z_i\}_{i=1}^n\in\ell_{\infty}[0,t_p)$, and meanwhile converge to zero as $t$ tends to $t_p$. Using the property of the hyperbolic-tangent-like function, we can proceed to prove that $\Gamma(\mathbf{z})\in\ell_{\infty}[0,t_p)$ and the convergence of which to zero at the prescribed time.
	
	By means of the auxiliary vectors as introduced in (\ref{auxiliaryvector}), one can find that $x_1=z_1$ and the following    closed-loop $z_1$-dynamics holds
	\begin{equation}\label{hengdengshi}
	\dot z_1=-k\mu h(z_1)+z_2
	\end{equation}
	where $z_2$ is a bounded function, which also can be  treated as a vanishing disturbance.  When $t\rightarrow t_p$, the  equivalent form of (\ref{hengdengshi}) is
	\begin{equation} \label{29}
	\dot x_1=\frac{-k}{t_p-t}\frac{e^{a x_1}-e^{-b x_1}}{a e^{a x_1}+b e^{-b x_1}}
	\end{equation}
	It follows from (\ref{Lemma11})-(\ref{x}) and (\ref{29}) that 
	\begin{equation}\label{199}
	e^{a x_1}-e^{-b x_1}= C_2{(t_p-t)^k} 
	\end{equation} 
	where $C_2=\left(e^{  a x_1(0) }-e^{-b x_1(0) }\right)/t_p^k$. Then taking time derivative on both sides of (\ref{199}), we have
	\begin{equation}\label{dhengdengshi}
	\begin{aligned}
	& {\left(a e^{a  x_1 }+b e^{-b  x_1 }\right) } x_2=kC_1(t_p-t)^{k-1},~k>n.\\
	\end{aligned}
	\end{equation}
	Observe that (\ref{dhengdengshi}) means that $x_2\rightarrow 0$ as $t\rightarrow t_p$. {Continue}, using the analysis similar to that used in (\ref{199})-(\ref{dhengdengshi}), by taking the $i$-th $(i=2,\cdots,n)$ derivative of both sides of (\ref{199}), we can generalize that $\{x_{i}\}_{i=2}^{n} $ and $\{\dot x_{i}\}_{i=1}^{n}$ converge to zero as $t\rightarrow t_p$ {(this is the reason for $k>n$)}.
	Therefore, $\lim_{t\rightarrow t_p}\|\mathbf{x}(t)\|=0$ and $\lim_{t\rightarrow t_p}\|\dot{\mathbf{x}}(t)\|=0$ hold. 
	
	In addition,  it follows from (\ref{auxiliaryvector}) that $J_n^{\top}\dot \Phi=\dot{\mathbf{z}}-\dot{\mathbf{x}}$, therefore $ \Phi=\dot{\mathbf{z}}-\dot{\mathbf{x}}+J_n^{\top}\mathbf{z}+\Gamma(\mathbf{z}) $,
	then
	\begin{equation}
	\begin{aligned}
	\|\Phi\|\leq\|\dot{\mathbf{z}}\|+\|\dot{\mathbf{x}}\|+\|\mathbf{z}\|+\|\Gamma(\mathbf{z})\|\in\ell_{\infty}[0,t_p).
	\end{aligned}
	\end{equation} 
	Consequently, from (\ref{u}) we have 
	\begin{equation}\label{999}
	\begin{aligned}
	|u(t)|&=|-B^{\top}(\mathcal{L}_n(\mathbf{a})\mathbf{x}+\Phi)|\\
	&\leq\|\mathcal{L}_n(\mathbf{a})\|\|\mathbf{x}\|+\|\Phi\| \\
	&\leq\|\mathcal{L}_n(\mathbf{a})\|\|\mathbf{x}\|+\|\dot{\mathbf{z}}\|+\|\dot{\mathbf{x}}\|+\|\mathbf{z}\|+\|\Gamma(\mathbf{z})\|.
	\end{aligned}
	\end{equation}
	Note that each term  in the third line of (\ref{999}) is bounded on $[0,t_p)$ and converges to zero as $t\rightarrow t_p$. Therefore,  $u(t)\in\ell_{\infty}[0,t_p)$ and $\lim_{t\rightarrow t_p}|u(t)|=0.$ This completes the proof.
	$\hfill\blacksquare$

 \begin{remark}
			When we consider a scalar system $\dot x=u$, from Theorem 1, one can immediately obtain a prescribed-time controller as $u_p=-\frac{k}{t_p-t}x$. Note that according to Theorem 1, this controller is only a special case under the design parameters $a$ and $b$ are chosen as small enough, and the design parameter $k$ satisfies $k>1$. Note that the classical finite-time controllers $u_f=-k\text{sign}(x)|x|^{\alpha}$  $(k>0,0<\alpha<1)$ (see \cite{Bhat 1998,Zhou 20}), then 
			the unique dynamic solution is 
			\begin{equation*}
			x(t)=\left\{\begin{array}{l}
			\text{sign}(x_0)\left(|x_0|^{1-\alpha}-k(1-\alpha)t\right)^{\frac{1}{1-\alpha}},~t\in[0,t_p)\\ 
			0,~t\geq t_p
			\end{array}\right.
			\end{equation*}
			where $t_p=\frac{|x_0|^{1-\alpha}}{k(1-\alpha)}$.
			Thereafter, the finite-time controller is equivalent to
			\begin{equation}\label{25}
			\begin{aligned}
			u_f=& -k\text{sign}(x)|x|^{\alpha-1}|x| =-k |x|^{\alpha-1} x \\
			=&-k\left(|x_0|^{1-\alpha}-k(1-\alpha)t\right)^{-1}x.
			\end{aligned}
			\end{equation}
			Inserting $|x_0|^{1-\alpha}=k(1-\alpha)t_p$ into  (\ref{25}), we have
			\begin{equation}\label{26}
			\begin{aligned}
			u_f &=-k\left(k(1-\alpha)t_p-k(1-\alpha)t\right)^{-1}x\\
			&=-\frac{x}{(1-\alpha)(t_p-t)},~  \forall t\in[0,t_p). 
			\end{aligned}
			\end{equation}
			Let $k=\frac{1}{1-\alpha}>1$, we have $u_p=u_f$. The above analysis   proves that the prescribed-time controller is equivalent to the finite-time controller under choosing some special design parameters.% but also inspires us to develop a new implementation scheme, that is, similar to  $\text{sign}(x)$ used in finite-time controller, denoting a function $\gamma(x)$ as
			%\begin{equation}\label{27}
			%\gamma(x)=\left\{\begin{array}{l}
		%	\frac{1}{t_p -t}\frac{x}{|x|}, x\neq 0\\ 
			%0,~~~~~~~x=0, 
			%\end{array}\right.
			%\end{equation}
			%Under such treatment,  the prescribed-time controller can be expressed as $u_p=-k\gamma(x)|x|,$ which   rejects the singularity since we have proven that $t=t_p$ is always accompanied by $x=0$.  
	\end{remark}
	
	\begin{remark}\label{remarknew}  
	 If we utilize  the PT-GUAS controller (as defined in (\ref{u}) with $k>n,~r=1$) and the  GUAS  controller ($k>0,~r=0$) through the following way, 
		\begin{equation*}
		u(t)=\left\{\begin{array}{l}
		- {B^{\top}} (\mathcal{L}_n(\mathbf{a})\mathbf{x}+J_n^{\top}(\dot{\Phi}+{\mathbf{z}})+(n+1)\mu H(\mathbf{z})),\\~~~~~~~~~~~~~~~~~~~~~~~~~~~~~~~~~~~~~~~~~~0\leq t<t_p\\
		- {B^{\top}} (\mathcal{L}_n(\mathbf{a})\mathbf{x}+J_n^{\top}(\dot{\Phi}+{\mathbf{z}})+  H(\mathbf{z})),~~t\geq t_p
		\end{array}\right.
		\end{equation*}	
		then the system states converge to zero as $t\rightarrow t_p$ and then remain zero for   $t\geq t_p$.  In fact, this switching method means that the prescribed-time controller guarantees that the closed-loop system is PT-GUAS on $[0,t_p)$ and the GUAS controller guarantees that the closed-loop system is ISS in the presence of some external disturbance on $[t_p,+\infty)$. 
	\end{remark} 
	%In particular, even if there are some mismatched and nonvanishing but bounded disturbances $d_*$ in each subsystem channel, the closed-loop system  belongs to input-to-state stable under such  switching treatment. For example, there exist a measurement error $d_{1*}$ in output state $x_1$, then it follows from (19) that the closed-loop $x_1$-dynamics becomes $\dot{x}_1=$
	
	\begin{remark}\label{remark3}  
		Various  methods on finite-time control have been reported in literature during the past few years, among which the most typical ones include  adding a power integral (AAPI), linear matrix inequalities (LMI) and implicit Lyapunov function (ILF), where the key element utilized is the fractional power state feedback (\textit{e.g.,} \cite{Amato2001,Poly 15,Sun 2017}).
		In pursuit of an alternative solution, we exploit a  unified control law  such that the  closed-loop system (\ref{dz}) can be regulated  asymptotically, exponentially or within prescribed-time by choosing the design parameters $k,~r$, $a$ and $b$ in (\ref{12}) properly. One salient feature with this method  is that it analytically  bridges the divide between prescribed-time control and traditional asymptotic control. Furthermore,  different design parameters ($a$ and $b$ in $h(x)$)    allow  different reshaped feedback signals to be utilized in the control scheme. Such treatment provides extra design  flexibility and degree of freedom in tuning regulation performance.
	\end{remark}

	\begin{remark}\label{remark4} 
		Compared with the existing prescribed-time control results  (see, for instance,  \cite{Song 17,Holloway 19a,Holloway 19b,Song 19}), 
		the proposed NTV feedback scheme, making use of the reshaped (compressed/expanded) feedback signal, is applicable over the entire operational process. In addition, this scheme has a numerical advantage over the aforementioned methods, this is because here only $1/(t_p-t)$ rather than   $1/(t_p-t)^n$ is involved for state scaling. Furthermore, with the proposed hyperbolic tangent function, the magnitude of initial control input can be adjusted through the parameters  $a$ and $b$. 
	\end{remark}
	%\begin{remark}\label{remark6} 
	%The proposed prescribed-time control involves recursive computation, gaining its motivation from  \textit{backstepping} design  ({\color{RoyalBlue}{\citealp{Krstic1995,Krstic19955}}}). Also, it is worth noting that the  algorithms here are connectively derived through one design step and one single Lyapunov function,  thus the design complexity does not increase with the order of the system.
	%\end{remark}	

	%%%%%%%%%%%%%%%%%%%%%%%%%%%%%%%%%%%%%%%%%%%%%%%%%%%%%%%%%%%%%%%%%%%%%%%%%%%%%%%%%%%%%
	
	\begin{example}
		To verify the effectiveness and benefits of control scheme as presented  Theorem 1, we conduct  a comparative simulation study through 	a third-order system 
		\begin{equation*}\label{3nd}
		\left\{\begin{array}{ll}
		\dot x_1(t)=x_2(t),\\
		\dot x_2(t)=x_3(t),\\
		\dot x_3(t)=u(t).
		\end{array}\right. 
		\end{equation*}
		According to Theorem 1-GUAS, the   asymptotically stabilization controller under $a,b\rightarrow 0$ is $u_{\text{GUAS}}=-K\mathbf{x}(t)$ with $K=[k^3+2k,3k^2+2,3k]^{\top}$ and $
		\mathbf{x}(t)=[x_1,x_2,x_3]^{\top},$  the control parameters are selected  as  $a=b=0.01$, $k=0.5$; and the initial condition are selected as $[x_1(0);x_2(0);x_3(0)]=[-1;0;1]$.
		According to Theorem 1-PT-GUAS, the following NTV feedback prescribed-time stabilization controller can be obtained, 
		\begin{equation}
		u_{\text{PT-GUAS}}= -z_2-\dot\phi_2-k\mu h(z_3),~t\in[0,t_p)
		\end{equation}
with		
		\begin{equation}\label{228} 
\begin{aligned}
	&	\mu(t,t_p)=1/(t_p-t),\\
	&	\phi_{1}(x_1,t)=k\mu h(x_1),\\
	&	\phi_{2}(x_1,x_2,t)=k\mu h(z_2)+x_1+\dot\phi_1,  \\
	&	h(z_i)=\frac{e^{a z_i}-e^{-b z_i}}{a e^{az_i}+\beta e^{-bz_i}},~i=1,2,3\\
	&	\dot\phi_i=\sum_{k=1}^{i}\frac{\partial \phi_i}{\partial x_k}x_{k+1}+\sum_{k=0}^{i-1}\frac{\partial \phi_i}{\partial\mu^{(k)}}\mu^{(k+1)},\\
	&	z_{1}=x_1,~z_2=x_2+\phi_{1},~z_3=x_3+\phi_{2}, 
\end{aligned}
		\end{equation}
		where   the corresponding design parameters are selected as  $a=b=1$, $k=6$, $t_p=6s$.   	In addition, to compare the control performance,   we adopt our previous result  (a linear feedback scheme in \cite{Song 17}) for simulation, the corresponding control law is given by $ 
		u_{S}=- \sum_{k=1}^{3}C_k^3 \frac{\mu_0^{(k)}}{\mu} x_{4-k}- v^{4}(k_1\dot{w}_1+k_2\ddot{w}_1)- k(\ddot{w}_1+k_1w_1+k_2\dot{w}_1)
		$, where $\mu_0=(t_p/(t_p-t))^4$, $v=t_p/(t_p-t)$, $w_1=\mu_0 x_1$.
		The design parameters in $u_{S}$ are selected as  $t_p=6.2s$, $k_1=k_2=0.6$ and $k=1$.

		\begin{figure}[htp]\label{fig4}
			\centering
			\includegraphics[width=3.4 in]{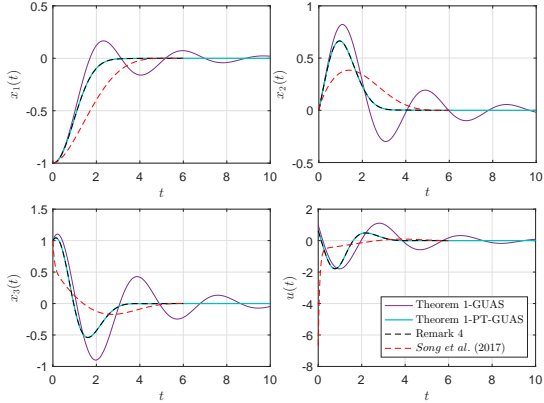} 
			\caption{The system state  and control input trajectories with $[x_1(0);x_2(0);x_3(0)]=[-1;0;1]$ and $t_p=6s$.}
		\end{figure}
		
		For comparison, simulation results obtained with the three different control schemes are shown in Fig. 2, from which asymptotic stabilization and prescribed-time stabilization are observed. Furthermore, it is seen that $i)$ the settling time with the proposed prescribed-time control is indeed irrespective of initial condition and any other design parameter;   $ii)$ the proposed scheme works within and after the prescribed-time interval; and $iii)$ compared with linear feedback scheme (black dotted line), it can be seen that the NTV feedback schemes (red dotted line and blue solid line) have a superior transient performance with a smaller initial control effort, verifying the effectiveness and benefits of the proposed algorithms.\footnote{Our design is another option in the control designer’s toolbox and we do not claim its universality with respect to the existing designs but highlight its better control performance of a class of LTI systems.} 
	\end{example}
	\subsection{Prescribed-time Observer}
	When only partial state is measurable, we employ  the prescribed-time observer proposed in  \cite{Holloway 19a}, to construct the prescribed-time control using output feedback.  As in  \cite{Holloway 19a} and \cite{Tian 17}, our solution is based on the separation principle, namely the controller is derived by designing a prescribed-time observer and a NTV output feedback control separately.

	{Considering that  $d(\mathbf{x}, t) \equiv 0$ and only output is available for feedback. The system (\ref{SISOsystem}) can be transformed into the following observer  canonical form  by  a linear nonsingular transformation $\xi(t)=M\mathbf{x}(t)$ 
		\begin{equation} \label{SISOsystem1}
		\left\{\begin{array}{ll}
		\begin{aligned}
		\dot{\mathbf{\xi}}(t)&=\mathscr{A}{\mathbf{\xi}}(t)+\mathscr{B}u(t)+\mathscr{D}y(t)\\
		y(t)&=  {\xi}_1(t)
		\end{aligned}
		\end{array}\right.
		\end{equation}
		where   $\mathscr{A}=J_n$, $\mathscr{B}=[b_{n-1},\cdots,b_0]^{\top}$,   $ \mathscr{D}=[a_{n },\cdots,a_1]^{\top}$, and the $a_i$s, $b_i$s are the same as those in (\ref{SISOsystem}).

		Invoking the  observer proposed in  \cite{Holloway 19a}, as follows
		\begin{equation}\label{observer}
		\begin{aligned}
		&\dot{\hat {\mathbf{\xi}}}(t)=\mathscr{A}{\hat{\mathbf{\xi}}}(t)+\mathscr{B}u(t)+\mathscr{D}y(t)\\
		&~~~~~~~~~~+\left[g_1(t,T),\cdots,g_n(t,T) \right]^{\top}(y-\hat{\xi}_1)
		\end{aligned}
		\end{equation}
		where  the time-varying observer gains $\{g_i(t,T)\}_{i=1}^n$   satisfy
		\begin{equation*}\label{gi}
		\begin{aligned}
		&g_i(t,T)=\left(\frac{n+m_0+i-1}{T}\bar{p}_{0_{i,1}}-\bar{p}_{0_{i+1,1}}\right)\mu_1^i\\
		&~~~~~~~~~~~~~-\sum_{j=1}^{i-1}g_j \bar{p}_{0_{i,j}}\mu_1^{n-j}+r_i,\\
		&g_n(t,T)=r_n+\frac{2n+m_0-1}{T}\bar{p}_{0_{n,1}}\mu_1^n -\sum_{j=1}^{n-1}g_j \bar{p}_{0_{n,j}}\mu_1^{n-j},
		\end{aligned}
		\end{equation*}
		where $\mu_1(t,T)=T/(T-t)$ and
		\begin{equation}\label{gn}
		\begin{aligned}
		&\bar{p}_{0_{i,i}}=1,~~~\bar{p}_{0_{i,j}}=0,~j\geq i\\
		&\bar{p}_{0_{i,j-1}}=-\frac{n+m_0+i-j}{T}\bar{p}_{0_{i,j}}+\bar{p}_{0_{i+1,j}},\\
		&~~~~~~~~~~~~~~~~~~~~~~~~~~~~~~~~~~  ~~~~~n-1\geq i\geq j\geq 2,\\
		&\bar{p}_{0_{n,j-1}}=-\frac{2n+m_0 -j}{T}\bar{p}_{0_{n,j}},~j=2,\cdots,n, 
		\end{aligned}
		\end{equation}
		and $m_0\geq1$ is an integer and $\mathbf{r}=[r_1,\cdots,r_n]^{\top}$ is selected to make the $n$-dimensional   matrix $\Lambda=\left[\mathbf{r},\left[I_{n-1},0_{1\times(n-1)}\right]^{\top}\right]$ Hurwitz. With (\ref{SISOsystem1})-(\ref{observer}) and observer error state $\tilde{\xi}=\xi-\hat{\xi}$,
		we get the observer error dynamics
		\begin{equation}\label{errordynamic}
		\begin{aligned}
		\dot{\tilde {\mathbf{\xi}}}(t)=&J_n{\tilde{\mathbf{\xi}}}(t)-\left[g_1(t,T),\cdots,g_n(t,T) \right]^{\top}\tilde{\xi}_1.
		\end{aligned}
		\end{equation}
		\begin{lemma}\label{lemma1}
			 {\cite{Holloway 19a}}  For the dynamic  system (\ref{SISOsystem}), consider the observer (\ref{observer}) having error dynamic (\ref{errordynamic}) and observer gains $\{g_i(t,T)\}_{i=1}^n$, and the $\{r_i\}_{i=1}^n$ are constants to be selected  such that the companion matrix $\Lambda$ is Hurwitz, then the closed-loop observer error system (\ref{errordynamic}) is prescribed-time stable, and there exist two positive constants $c_1$ and $c_2$ such that
			\begin{equation}\label{D-tildex}
			|\tilde{\mathbf{\xi}}(t)|\leq\mu_1(t,T)^{-m_0-1}c_1\exp(-c_2 t)|\tilde {\mathbf{\xi}}(0)|,
			\end{equation}
			for all $t\in[0,T)$. In addition, the output estimation error injection terms $\{g_i(t,T)\tilde {\xi}_i\}_{i=1}^{n}$  remain uniformly bounded over $[0,T)$, and  converge to zero as $t\rightarrow T.$ Also, ${\tilde{\mathbf{x}}}(t) $ has the same dynamic properties as ${\tilde{\mathbf{\xi}}}(t)$ since ${\tilde{\mathbf{x}}}(t)=M^{-1}{\tilde{\mathbf{\xi}}}(t)$ with $M$ being a nonsingular constant matrix.
		\end{lemma}
		
		\subsection{Prescribed-time Output Feedback Controller} 			
		{The output feedback prescribed-time control law for  system (\ref{SISOsystem}) is constructed by replacing   $x_1,x_2,\cdots,x_n$ with $\hat{x}_1,\hat{x}_2,\cdots,\hat{x}_n$ in  (\ref{u}) as follows:}
		\begin{equation}\label{51}
		u=- {B^{\top}} \left(\mathcal{L}_n(\mathbf{a})\hat{\mathbf{x}}+ {\Phi}(\hat{\mathbf{x}})\right)
		\end{equation}
		where only $x_1$ is measurable, ${\Phi}(\hat{\mathbf{x}})=J_n^{\top}(\dot\Phi(\hat{\mathbf{x}})+\hat{\mathbf{z}})+\Gamma(\hat{\mathbf{z}})$ with $\Gamma(\hat{\mathbf{z}})=k\mu^r[h_1(\hat{z}_1),\cdots,h_n(\hat{z}_n)]^{\top}$ and $\hat{\mathbf{z}}=\hat{\mathbf{x}}+J_n^{\top}\Phi(\hat{\mathbf{x}}).$ 
		
		The control law (\ref{51})  involves the design parameters $k$ and $r$ as defined  in Theorem 1. It can be verified that different $k$ and $r$ lead to different convergence rate. For instance, we can set $k > r$  and $r= 0$, and by invoking the classical high-gain observer, to achieve asymptotic or exponential output regulation. Here in this subsection, our ambitious goal is to achieve state regulation with output feedback with the aid of the prescribed-time observer developed in \cite{Holloway 19a}.

		\begin{theorem}
			{For the dynamic system (\ref{SISOsystem}) with $d(\mathbf{x},t)\equiv\mathbf{0}$, consider the output feedback control law (\ref{51})  with   the prescribed-time observer (\ref{observer}),  the closed-loop system is prescribed-time stable if the controller and the observer parameters are selected according to Theorem 1-PT-GUAS and Lemma 2,   and $T\leq t_p$.}
		\end{theorem}
	 
		\textit{Proof:} The proof consists of  two steps, the first step is to  prove that the closed-loop system with the observer and the output feedback control scheme does not escape during $[0,T)$, and the second step is to show that all closed-loop trajectories converge to  zero as $t$ tends to $t_p$ and remain zero thereafter.

		\textit{Step 1:} We consider the Lyapunov function $V=\mathbf{z}^{\top}\mathbf{z}/2$. Using $J_nJ_n^{\top}+BB^{\top}=I_n$ and $BB^{\top}\mathcal{L}_n(\mathbf{a})=\mathcal{L}_n(\mathbf{a})$, the derivative of $V$ over $[0,T)$ along (\ref{SISOsystem}) under the output feedback control law (\ref{51}) becomes 
		\begin{equation*}
		\begin{aligned}
		\dot V
		&=\mathbf{z}^{\top}\left[\left(J_n+\mathcal{L}_n(\mathbf{a})\right){\mathbf{z}}-J_nJ_n^{\top}{\Phi( {\mathbf{x}})}-\mathcal{L}_n(\mathbf{a})J_n^{\top}{\Phi( {\mathbf{x}})}\right]\\
		&~~~-\mathbf{z}^{\top}\left[BB^{\top}\mathcal{L}_n(\mathbf{a})(\mathbf{x}-\tilde{\mathbf{x}})+BB^{\top}{(\Phi(\mathbf{x})-\Phi(\tilde{\mathbf{x}}))}\right]\\
		&~~~+\mathbf{z}^{\top}\left[{\Phi( {\mathbf{x}})}-\Gamma(\mathbf{z})-J_n^{\top}{\mathbf{z}}\right] \\
		&=\mathbf{z}^{\top}\left(J_n+\mathcal{L}_n(\mathbf{a})-J_n^{\top} \right){\mathbf{z}}-\mathbf{z}^{\top}{\Phi( {\mathbf{x}})}+\mathbf{z}^{\top}BB^{\top}{\Phi(\tilde{\mathbf{x}})}\\
		&~~~-\mathbf{z}^{\top}\mathcal{L}_n(\mathbf{a})\left(\mathbf{z}-\tilde{\mathbf{x}}\right) +\mathbf{z}^{\top}\left({\Phi( {\mathbf{x}})}-\Gamma(\mathbf{z})\right)\\
		&=-\mathbf{z}^{\top}\Gamma(\mathbf{z})+\mathbf{z}^{\top}\left(\mathcal{L}_n(\mathbf{a})\tilde {\mathbf{x}}+BB^{\top}{\Phi(\tilde{\mathbf{x}})}\right)
		\end{aligned}
		\end{equation*}
		where $\tilde {\mathbf{x}}={\mathbf{x}}-\hat{\mathbf{x}}$, ${\Phi(\tilde{\mathbf{x}})}= {\Phi}( {\mathbf{x}})-{\Phi(\hat{\mathbf{x}})}$. It is seen from Lemma \ref{lemma1}  that $\tilde{\mathbf{x}}$ remains uniformly bounded over $[0,T)$ and converges to zero as $t\rightarrow T$. The boundedness of ${\Phi(\tilde{\mathbf{x}})}$ is also guaranteed by the bounded $\tilde{\mathbf{x}}$, and ${\Phi(\tilde{\mathbf{x}})}$ also converges to zero as $t$ tends to $T$. Therefore,   there exist a positive constant $\gamma<\infty$ such that $\dot V\leq \gamma$ holds for $\forall t\in[0,T)$. It follows that $V$ and $\mathbf{z}$  cannot escape during the interval $[0,T).$ 
		
		\textit{Step 2:} From Lemma \ref{lemma1}, we know that there exists a prescribed-time $T$, such that $\{\hat{x}_i(t)=x_i(t)\}_{i=2}^n$ for $t\geq T$. In consequence, the output feedback control law $u=- {B^{\top}}\left(\mathcal{L}_n(\mathbf{a})\hat{\mathbf{x}}+ {\Phi}(\hat{\mathbf{x}})\right)$ coincides with state feedback control law $u=-B^{\top}\left(\mathcal{L}_n(\mathbf{a}) {\mathbf{x}}+ {\Phi}( {\mathbf{x}})\right)$ for  $\forall t\geq T$. In other words,  this output feedback law can be used to establish prescribed-time stability and performance recovery (see  \cite{Khalil(2002),Tian 17} and  \cite{Holloway 19b}). Note that the closed-loop trajectory under (\ref{51}) does not escape during  $t\in[0,T)$, it follows from Theorem 1-PT-GUAS that under the proposed output feedback control law, there exists another pre-set time $t_p\geq T$ to steer the system from an arbitrary bounded state to zero as $t\rightarrow t_p.$  The boundedness of $u(t)$ can also be easy established according to Theorem 1. This completes the proof. $\hfill\blacksquare$

		\begin{remark} To close this section, it is worth making the following comments.
			\begin{itemize}
				\item[$i)$] The output feedback controller inherits the properties and advantages of the  state feedback controller  as stated in Theorem 1, that is, elegant parameter tuning,  one-step design process, simple controller structure. 
				\item[$ii)$] Prescribed-time  stabilization is the result of employing the scaling function $\mu(t,t_p)$ and the nonlinear feedback function $h(x)$  inside the control  {scheme} (\ref{u}). All observer errors $\{\tilde{x}_i\}_{i=1}^n$ are regulated to zero  {within the} pre-set time $T $, and all system states $\{{x}_i\}_{i=1}^n$ are regulated to zero  {within} pre-set time $t_p $, where  $t_p\geq T$. 
				\item[$iii)$]  
				{Only the output state $(x_1)$ is required in constructing the output feedback prescribed-time control (\ref{51}).  Such control scheme   has  an obvious numerical advantage  because its maximum implemented gain is $1/(t_p-t)^n$, while the standard linear $n$-th order output feedback prescribed-time controller proposed in  \cite{Holloway 19b}  involves $1/(t_p-t)^{2n}$.}
				\item[$iv)$]  It is noted that the time-varying gains  $\mu(t,t_p)$, $\mu_1(t,T)$ go to infinity as time tends to $t_p$ or $T$, such phenomenon (unbounded control gain at the fixed convergence time) appears  in all results on strictly finite-/fixed-/prescribed-time control. The implementation solution  for finite-/fixed-time control is to use fractional power state feedback or sign functions switching the controller to zero when the system is regulated to the equilibrium point. The other implementation solutions for prescribed-time control are given in  \cite{Holloway 19b,Krishnamurthy 20a,Song 17}.  One typical way to address this issue is to let the system operate  in a finite-time interval, \textit{i.e.,} adjusting the operational time slightly shorter than prescribed convergence time.   Another typical way is to  let  the system operate in an infinite time interval by making suitable saturation on the control gains. Here in this work,  we use the method as described  in Remark 3 to implement the  prescribed-time controller for $t\in[0,t_p)$ and initiate the asymptotically stable controller for $t\in[t_p,+\infty)$. 
			\end{itemize}
		\end{remark}

		\begin{example}  We illustrate the performance of the observer  and the output feedback controller through the following model:  
			\begin{equation*}
			\left\{\begin{array}{ll}
			\dot x_1(t)=x_2(t)~\\
			\dot x_2(t)=u(t)~\\
			y(t)=x_1(t),
			\end{array}\right. 
			\end{equation*} 
			the observer   is	
			\begin{equation}
			\left\{\begin{array}{ll}
			\dot{\hat{x}}_{1}(t)=\hat{x}_{2}(t)+g_{1}\left(t, T\right)\left(y(t)-\hat{x}_{1}(t)\right) \\
			\dot{\hat{x}}_{2}(t)=u(t)+g_{2}\left(t, T\right)\left(y(t)-\hat{x}_{1}(t)\right)
			\end{array}\right.
			\end{equation}
			with $g_1(t,T)=r_1+\frac{2(m_0+2)}{T}\mu_1$, $g_2(t,T)=r_2+r_1\frac{m_0+2}{T}\mu_1+\frac{(m_0+1)(m_0+2)}{T^2}\mu_1^2$.  For observer parameters, we select $r_1=r_2=1$, $m_0=3$, $T=4$s and $[\hat x_1(0);\hat x_2(0)]=[0;0]$.	For output feedback controller parameters, we select $k=5$, $a=b=1$ and $t_p=6$s. The initial condition is $[x_1(0);x_2(0)]=[4,-3]$.  The control law $u$  is implemented similar to (\ref{228}), just replacing $ x_2$ in (\ref{228}) with $ \hat x_2$. 
			Furthermore, the control performance in a noisy environment is studied by considering the output signal $y(t)$  corrupted with an uncertain measurement  noise $\eta(t)$,  namely $y(t)=x_1(t)+\eta(t)$ with $\eta(t)=0.001\sin(3t)$.  
			The closed-loop state $\{x_i(t)\}_{i=1}^2$ trajectories,  state estimate $\{\hat x_i(t)\}_{i=1}^2$ trajectories, the norm of observer estimation error $\|\tilde{\mathbf{x}}(t)\|$, the norm of system state $\| {\mathbf{x}}(t)\|$ and control input signal $u(t)$ are shown in Figs. 3-4. 
			
			It is observed from Fig. 3  that the controller remains operational after $t_p$, and all closed-loop signals are bounded on the whole time-domain, in particular,
			the observer estimation errors converge  to zero  as $t\rightarrow T(4s)$, and system states converge to zero   as $t\rightarrow t_p(6s)$,  confirming our theoretical prediction and analysis.  Fig. 4 shows that the proposed controller retains its performance %(\textit{i.e.,} both the state and state estimate converge to a small neighborhood of the origin in the prescribed time)  
			even in the presence of  measurement  noise. Although a slightly chattering phenomenon, caused by noise and controller switching, occurs near $t_p$, the control input remains bounded on the whole time-domain. In addition, the numerical advantage leading to friendly implementation  has been verified in simulation.
			\begin{figure}[t] \label{fig5}
				\centering
				\includegraphics[height=3.7cm]{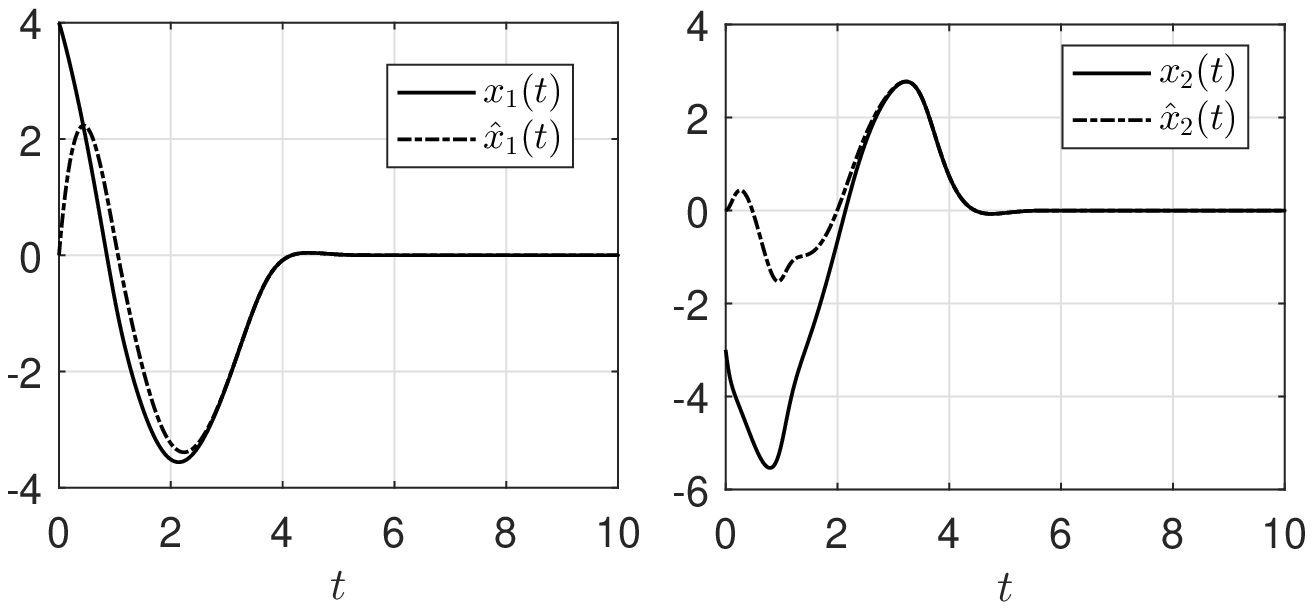}\\
				\includegraphics[height=3.7cm]{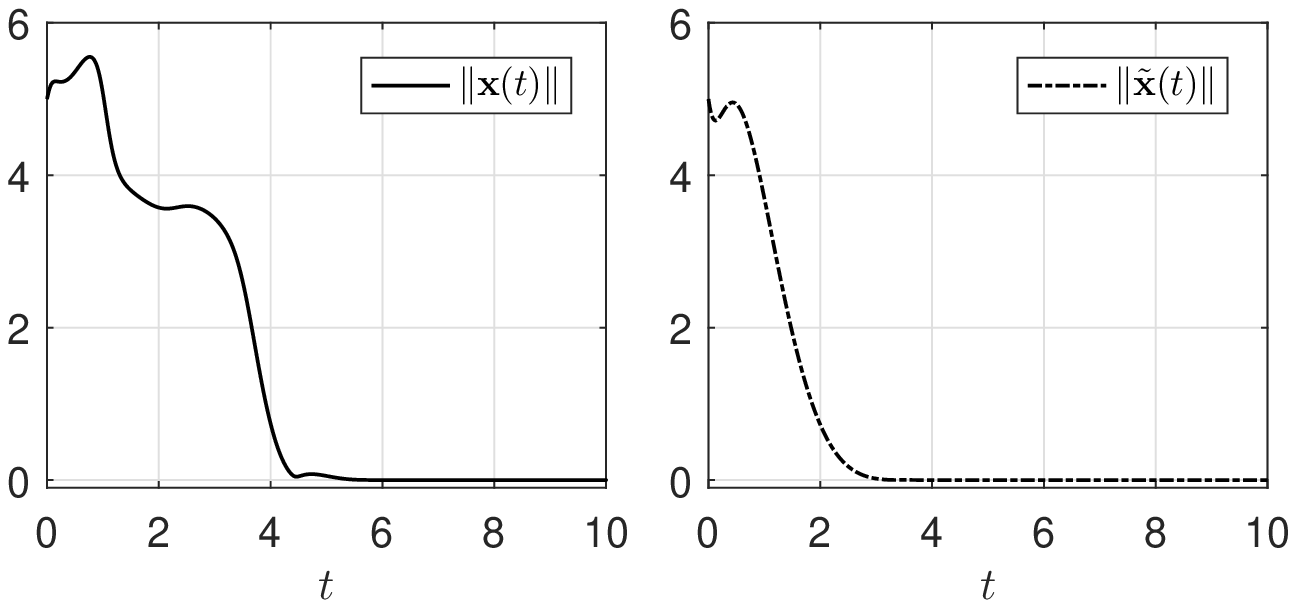}\\
				\includegraphics[height=3.3 cm]{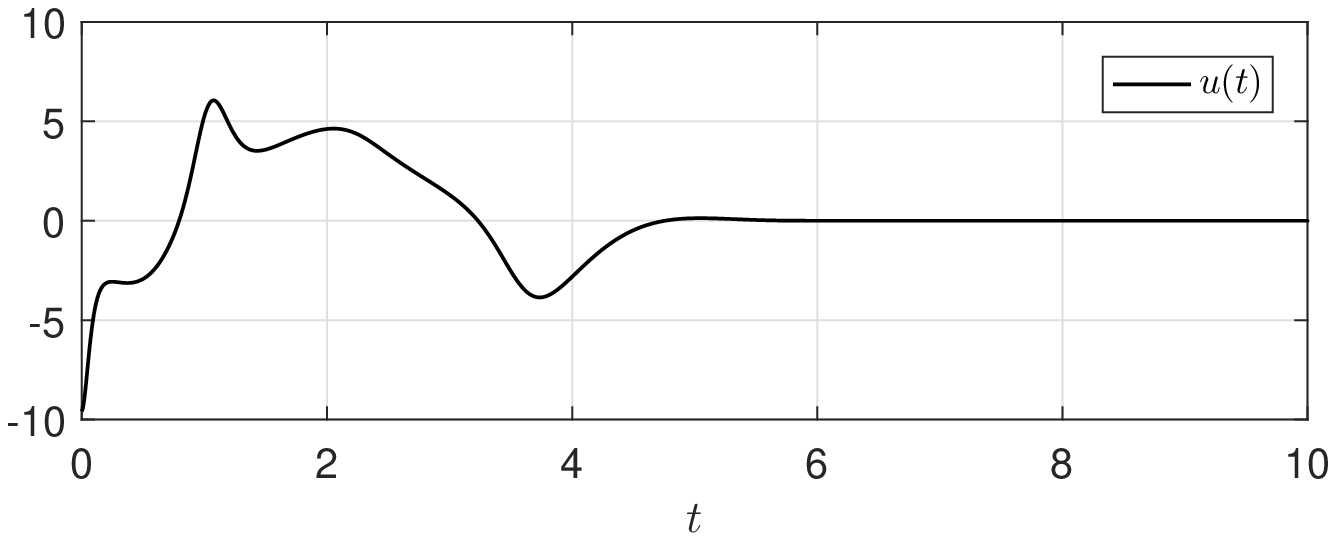} 
				\caption{Simulation results with prescribed-time observer ($T=4s$) and the proposed output feedback control ($t_p=6s$).}
			\end{figure}
			\begin{figure}[t] \label{fig6}
				\centering
				\includegraphics[height=3.7cm]{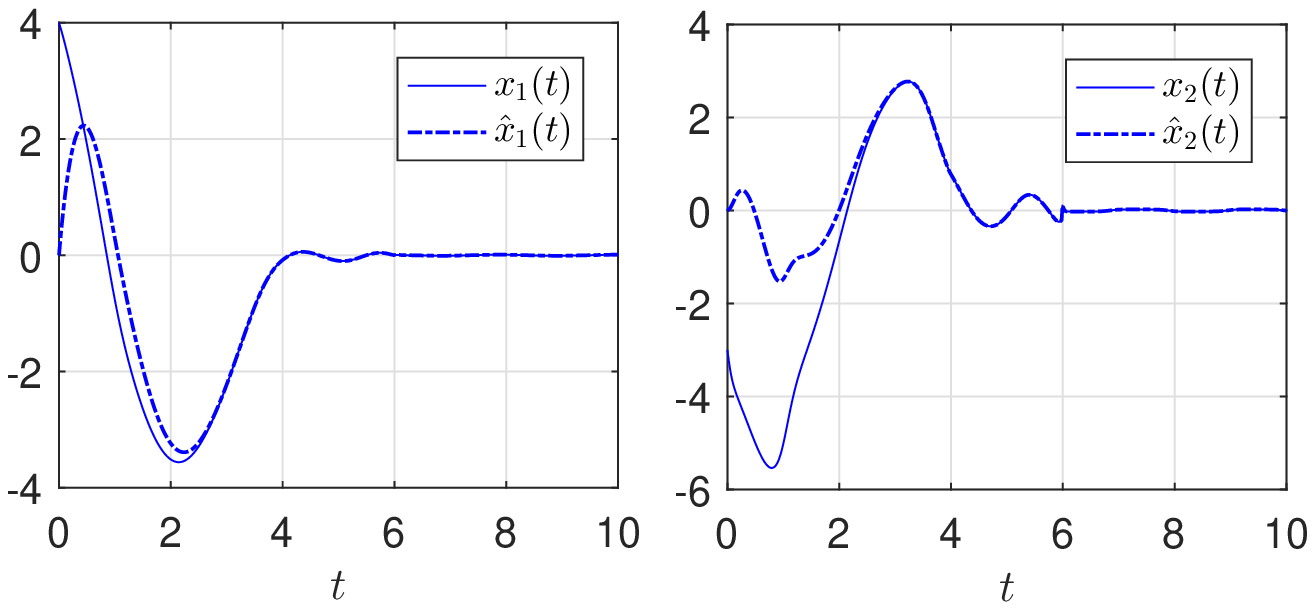}\\
				\includegraphics[height=3.7cm]{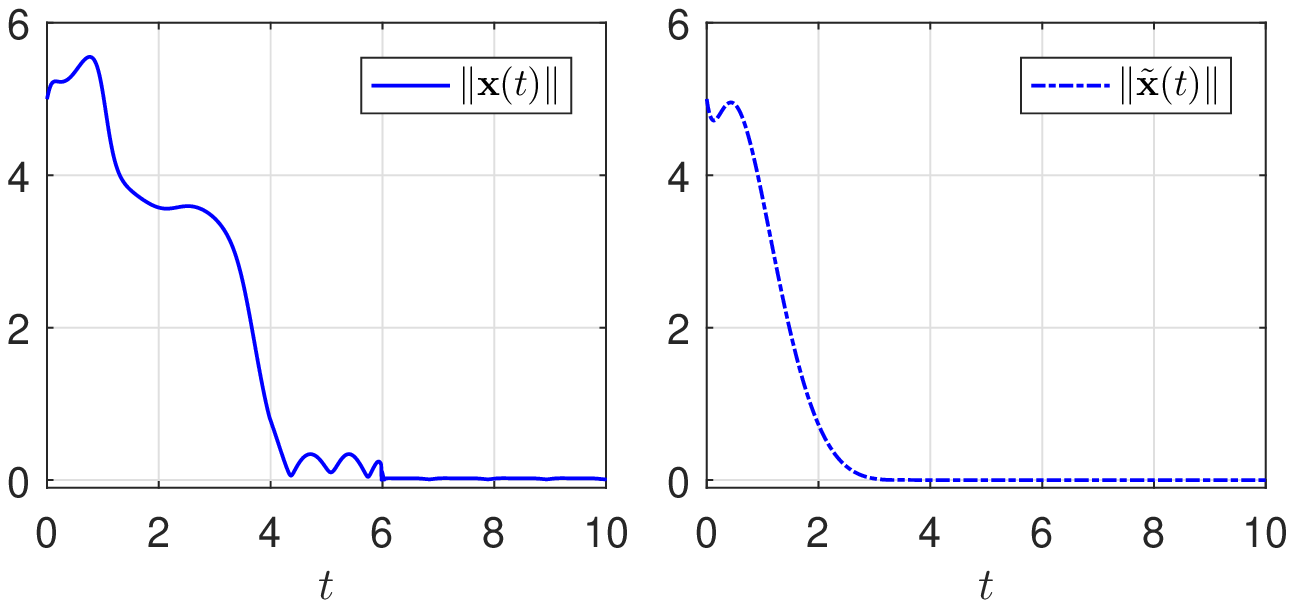}\\
				\includegraphics[height=3.3cm]{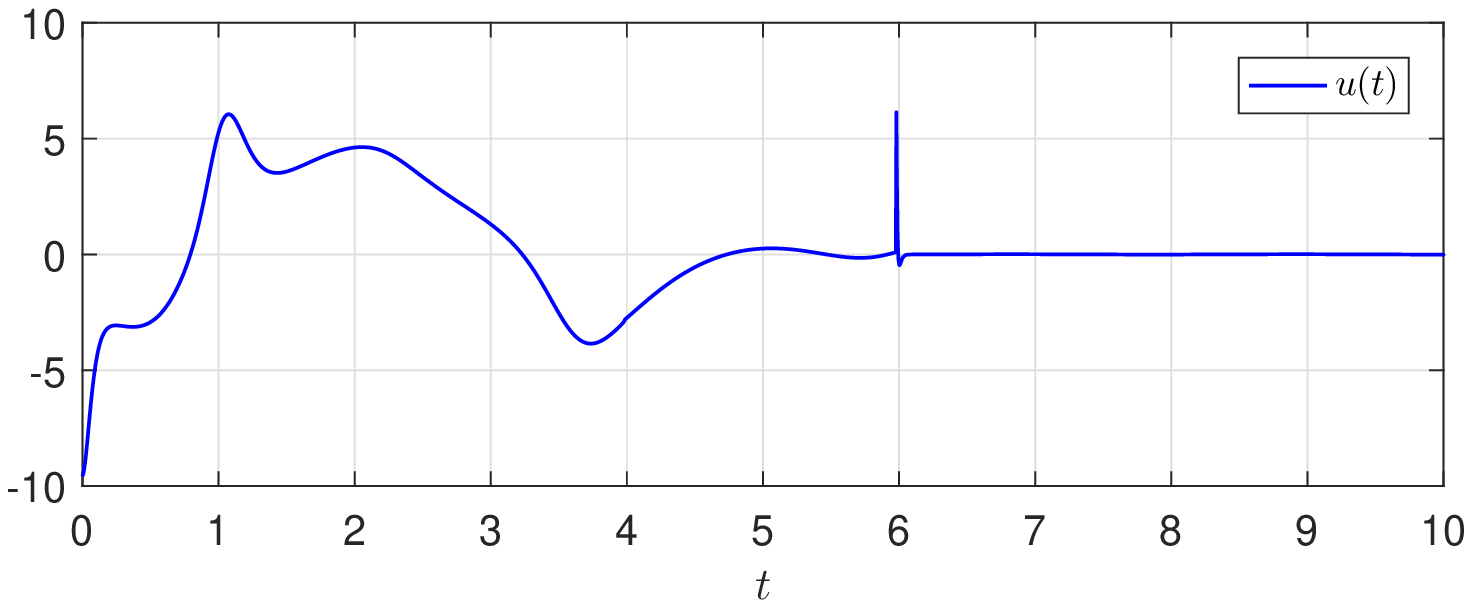} 
				\caption{Simulations results with measurement noise.}
			\end{figure} 
		\end{example}
		\section{Prescribed-time control for  linear systems in canonical form  with  uncertainties}\label{section4}
		In the presence of non-vanishing uncertain term  $d(\mathbf{x},t)$,   system (\ref{SISOsystem}) can be rewritten as
		\begin{equation}\label{1t}
		\left\{\begin{array}{ll}
		\dot x_i=x_{i+1},~~~~i=1,\cdots,n-1\\
		\dot x_n= u+F(x_1,\cdots,x_n,t)
		\end{array}\right.
		\end{equation} 
		where $F(\cdot)=\sum_{i=1}^{n}a_{i}x_i+d(\mathbf{x},t) $ is an unknown smooth function  and  satisfies $|F(\cdot)|\leq \bar{d} (\mathbf{x})$ with $\bar{d}(\cdot)$   being a known scalar real-valued function.
		
		Define a sliding surface $s(t)$ on $[0,t_p)$  as follows:
		\begin{equation}\label{SS}
		s(t)= 
		x_n+\phi_{n-1}(x_1,\cdots,x_{n-1},t),~~~~~~~~~~~~
		\end{equation} 
		where   $\phi_{n-1}$ is defined in (\ref{auxiliaryvector}). Some other sliding surface selection can be referred to  \cite{Harl} and \cite{Chen 2021}.		
		The derivative of the auxiliary variable along the trajectories of (\ref{1t}) is% and the element $h(z_{n-1})$ in $\phi_{n-1} $ is defined in (\ref{hx}). %and $\{l_i\}_{i=1}^{n-1}$   are assigned such that the polynomial $l_1+l_2s+\cdots+l_{n-1}s^{n-2}+s^{n-1}$ is Hurwitz. 
		\begin{equation}\label{ds}
		\dot s=u+F(\cdot)+\dot\phi_{n-1}.
		\end{equation}
		where $\dot\phi_{n-1}=\sum_{k=1}^{n-1}\frac{\partial \phi_{n-1}}{\partial x_k}x_{k+1}+\sum_{k=0}^{n-2}\frac{\partial \phi_{n-1}}{\partial \mu^{(k)}}\mu^{(k+1)}$ belongs to a computable function. 
		\begin{theorem}
			Consider system (\ref{SISOsystem}) and the transformed system (\ref{1t}), the closed-loop signals $\{x_i\}_{i=1}^{n}$ and  $s(t)$ are prescribed-time globally uniformly asymptotically stable (PT-GUAS),  if the control  law is designed as: 
			\begin{equation}\label{PTSMC}
			u= 
			-\bar{d} \text{sign}(s)-\dot\phi_{n-1}-k\mu h(s), 
			\end{equation}
			where $k>n$, $\bar d\geq F(\cdot)$, $\mu=1/(t_p-t)$, $\dot\phi_{n-1}$ is a computable function as described after (\ref{ds}), and $h(\cdot)$ is the hyperbolic-tangent-like function as defined in (\ref{hx}).  
		\end{theorem}
	
		\textit{Proof:}  For $t\in[0,t_p)$, let $V=|s|$. With the control scheme  (\ref{PTSMC}), for $s\neq 0$, the upper right-hand derivative of $V$ along the trajectory of the closed-loop system (\ref{1t}) becomes 
		\begin{equation}\label{45}
		D^* V=\frac{s\dot s}{|s|}=-(\bar d -F)-k\mu h(|s|)\leq-k\mu h(V).
		\end{equation}
		By using Lemma 1,   it is easy to get that    $V\in \ell_{\infty}[0,t_p)$  and $\lim_{t\rightarrow t_p}V=0$, establishing the same for $s(t)$ and $\dot s(t)$. At the same time, the closed-loop $\{x_i\}_{i=1}^{n-1}$-dynamics become
		\begin{equation}
		\begin{aligned}
		\dot x_i=&x_{i+1};~~~~i=1,\cdots,n-2\\
		\dot x_{n-1}=&-\phi_{n-1}
		\end{aligned}
		\end{equation}
		where $\phi_{n-1}$ is the virtual control input. It is seen that  the  control law (\ref{PTSMC})  reduces the perturbed $n$-th order system to an unperturbed $(n-1)$-th order system.  Therefore, by using Theorem 1-PT-GUAS, we can prove that the closed-loop signals $\{x_i\}_{i=1}^{n-1}$, $\{\dot x_i\}_{i=1}^{n-1}$, $\phi_{n-1}$ and $\dot\phi_{n-1}$ are bounded and converge to zero as $t\rightarrow t_p$. From (\ref{45}) and the analysis process in Section 3, it is not difficult to verify that $\mu h(s)$ is also bounded. Therefore, it follows from (\ref{PTSMC})  that the control input $u(t)$ is bounded for $t\in[0,t_p)$. This completes the proof.  $\hfill\blacksquare$ 
		
		\begin{remark}\label{remark6}  
			For $t\in[t_p,+\infty)$, we specifically design $s(t)$ as $s(t)=x_n+\sum_{i=1}^{n-1}l_ix_{i} $, where $\{l_i\}_{i=1}^{n-1}$ are assigned such that the polynomial $l_1+l_2s+\cdots+l_{n-1}s^{n-2}+s^{n-1}$ is Hurwitz, and design the corresponding control law as $u=-\bar d\text{sign}(s)-\sum_{i=1}^{n-1}l_ix_{i+1}$. As a result, $D^*|s(t)|\leq 0$, we therefore obtain that $s(t)=0,\forall t\in[t_p,+\infty)$ by recalling that $s(t_p)=0$.  Furthermore, it is not difficult to get  $u\in \ell_{\infty}[t_p,+\infty)$.  
			As the disturbances do not disappear,  the control action  for $t\geq t_p$  is no longer  zero but bounded,  a necessary effort to fight against the ever-lasting (nonvanishing) uncertainties/disturbances,  which is comprehensible in order to maintain each state at the equilibrium (zero)  after the prescribed settling time. 
		\end{remark}

		\begin{example}
			To verify the effectiveness of the  prescribed-time sliding mode controller, we consider the following system:
			\begin{equation*} \label{dsimulation2nd}
			\left\{\begin{array}{ll}
			\dot x_1(t) =x_2(t) ,~\\
			\dot x_2(t) =u (t)+F(x_1,x_2,t) 
			\end{array}\right. 
			\end{equation*}
			 where  \begin{equation}
			 F(x_1,x_2,t)=0.03x_1+0.01\sin(x_2)+0.02\sin(2t)
			 \end{equation}  				
			here $\bar d$  can be selected as $\bar d= 0.03(|x_1|+1)$. According to Theorem 3, the controller is given by
			\begin{equation}
							u= -\bar d \text{sign}(s)-\dot\phi_1-k\mu h(s),~t\in[0,t_p) 
			\end{equation}
			with
			\begin{equation*}\label{229} 
\begin{aligned}
		&	\phi_{1}=k\mu h(x_1),  ~~
			\mu=1/(t_p-t),\\
		&	\dot\phi_1= \frac{\partial \phi_1}{\partial x_1}x_{2}+ \frac{\partial \phi_1}{\partial\mu}\dot\mu,\\
		&	h(x)=(e^{a x}-e^{-b x})/(a e^{a x}+b e^{-bx}), \\
		&	s= 
			x_2+\phi_1,~~~~t\in[0,t_p)  \\
		&	u= -\bar d \text{sign}(s)-\dot\phi_1-k\mu h(s),~t\in[0,t_p) 
\end{aligned} 
			\end{equation*} 
			In addition, according to Remark 6, we design $u=-\bar d \text{sign}(s)-l_1x_2, ~\forall t\geq t_p$ with $s=x_2+l_1x_1$. 
			For simulation,   the design parameters are chosen as $a=b=l_1=1$ and $k=3$. To verify the property of prescribed-time convergence  \textit{w.r.t.} the initial conditions, three different initial values $[x_1(0);x_2(0)]=[ 1;-1]$, $[x_1(0);x_2(0)]=[2;-1]$ and $[x_1(0);x_2(0)]=[3;-1]$   are considered in Fig. 5. To confirm the property of prescribed-time convergence \textit{w.r.t.} $t_p$, we choose $t_p=2s,3s,4s$ respectively in Fig. 6.

			Simulation results show that: $i)$ all states converge to zero synchronously within the pre-set time $t_p$; and $ii)$ the convergence time is independent of initial conditions and any other design parameter. Certain control  chattering is observed as $t\rightarrow t_p$,  which is caused by the auxiliary variable $s$ and controller switching. Specially, the magnitude of control input  has a slight increase when $|x_1(0)|$  increases or $t_p$ decreases.  
						\begin{figure}[htp]\label{fig9}
				\includegraphics[height=4.2cm]{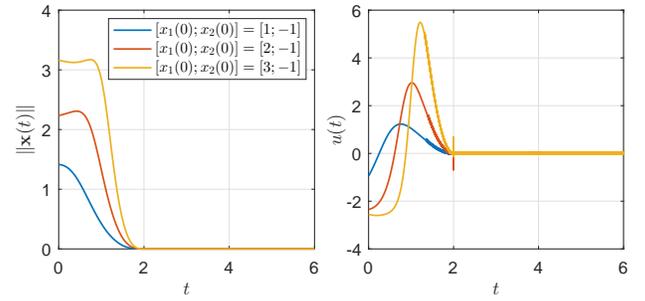} 
				\caption{Simulation results   with different initial conditions under  $ t_p=2s$.}
			\end{figure}
			\begin{figure}[htp]\label{fig8}
				\includegraphics[height=4.2cm]{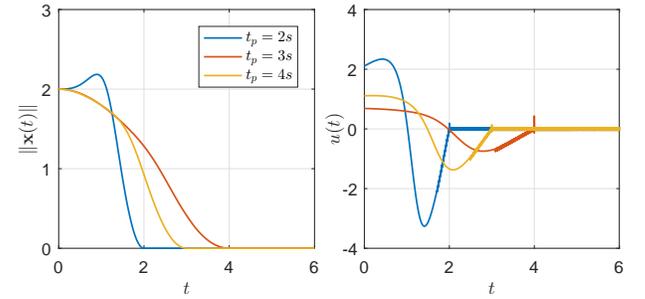}
				\caption{Simulation results   with different prescribed-time under   $[x_1(0);x_2(0)]=[-2;0]$.}
			\end{figure}
		\end{example}

		\section{Conclusions}\label{section5}
		A unified nonlinear and time-varying feedback control scheme is developed to achieve prescribed-time regulation of high-order uncertain systems.  The proposed control is able to achieve asymptotic, exponential, or prescribed-time  regulation by selecting the design parameters properly. The rule of parameter selection has been given through the Lyapunov theory. Furthermore,  prescribed-time output feedback control and prescribed-time sliding mode control for high-order systems are developed, where the advantages of simplicity (elegancy) yet superiority are retained.  				
		Extension of the proposed method to more general nonlinear systems with  mismatched uncertainties   represents an interesting future research topic.

\end{document}